# Designing transformation-induced plasticity and twinning-induced plasticity Cr-Co-Ni medium entropy alloys: theory and experiment

Zhibiao Yang[1,2], Song Lu[2,a], Yanzhong Tian[3,b], Zijian Gu[3], Huahai Mao[4,5], Jian Sun[1,c], Levente Vitos[2,6,7]

1 Shanghai Key Laboratory of Advanced High-Temperature Materials and Precision Forming, School of Materials Science and Engineering, Shanghai Jiao Tong University, Shanghai 200240, People's Republic of China
2 Applied Materials Physics, Department of Materials Science and Engineering, Royal Institute of Technology, Stockholm SE-100 44, Sweden
3 Key Laboratory for Anisotropy and Texture of Materials (Ministry of Education), School of Materials Science and Engineering, Northeastern University, Shenyang 110819, China
4 Unit of Structures, Dept Materials Science and Engineering, KTH, SE-10044, Stockholm, Sweden
5 Thermo-Calc Software, Råsundav. 18, SE-16767, Solna, Sweden
6 Division of Materials Theory, Department of Physics and Materials Science, Uppsala University, P.O. Box 516, Uppsala SE-75120, Sweden
7 Research Institute for Solid State Physics and Optics, Wigner Research Center for Physics, Budapest H-1525, P.O.Box 49,Hungary

## Abstract

In order to efficiently explore the nearly infinite composition space in multicomponent solid solution alloys, it is important to establish predictive design strategies and use computation-aided methods. In the present work, we demonstrated the density functional theory calculations informed design routes for realizing transformation-induced plasticity (TRIP) and twinning-induced plasticity (TWIP) in Cr-Co-Ni medium entropy alloys (MEAs). We systematically studied the effects of magnetism and chemical composition on the generalized stacking fault energy surface (γ-surface) and showed that both chemistry and the coupled magnetic state strongly affect the γ-surface, consequently, the primary deformation modes. Based on the calculated effective energy barriers for the competing deformation modes, we constructed composition and magnetism dependent deformation maps at both room and cryogenic temperatures. Accordingly, we proposed various design routes for achieving desired primary deformation modes in the ternary Cr-Co-Ni alloys. The deformation mechanisms predicted by our theoretical models are in nice agreement with available experimental observations in literature. Furthermore, we fabricated two non-equiatomic Cr-Co-Ni MEAs possessing the designed TWIP and TRIP effects, showing excellent combinations of tensile strength and ductility.

Corresponding authors:

aSong Lu, songlu@kth.se

bYanzhong Tian, tianyanzhong@mail.neu.edu.cn

cJian Sun, jsun@sjtu.edu.cn

# 1. Introduction

High entropy alloys (HEAs) were originally proposed by Yeh and Cantor in 2004 [1,2] as equiatomic concentrated solid solutions which stabilize the single-phase structure through the maximized configurational entropy. Based on this concept, many types of single-phase HEAs have been successfully developed and some of them have been demonstrated possessing excellent mechanical properties, including CrCoNi medium entropy alloy (MEA) [3,4], CrFeCoNi [5,6] and CrMnFeCoNi [2,7] HEAs. Besides the effect of mixing entropy, studies have focused on the chemical effect on the phase stability [8]. Recent development in HEAs has already weakened the restriction of entropy maximization, in order to explore the vast composition space in these multicomponent alloys. In particular, Li et al. [9] proposed the so-called 'metastability-engineering strategy' in Cr-Mn-Fe-Co alloys to optimize the mechanical performance. The composition of the alloy was designed to alter the thermodynamic stability of the fcc phase, in order to trigger the transformation-induced plasticity (TRIP) [9,10] or the twinning-induced plasticity (TWIP) effects [11,12]. This design concept is to combine the benefits from the massive solid solution hardening in HEAs with the TRIP/TWIP effects. Following this idea, they have designed TRIP or TWIP assisted single- and dual-phase HEAs with intriguing mechanical properties in Cr-Mn-Fe-Co alloys [9,12]. In fact, equiatomic MEAs/HEAs such as CrCoNi and CrMnFeCoNi are also strengthened by the TRIP and/or TWIP effects [3,4,7]. In particular, the CrCoNi MEA exhibits extraordinary combination of high strength, high ductility and high fracture toughness [3,4], superior to CrMnFeCoNi and any other investigated ternary or quaternary equiatomic variants. More interestingly, the mechanical properties of these HEAs are further improved at cryogenic temperatures, which are ascribed to the enhanced TRIP/TWIP effects via the occurrence of twin+hcp lamellar structure, partly due to the temperature effect on the stacking fault energy (SFE) [4].

The primary deformation modes in face-centered cubic (fcc) metals and alloys, namely, martensitic phase transformation (MT), twinning (TW), and dislocation slip (SL), depend strongly on the size of the SFE [13-15]. Empirically, their relationship has been well documented in many alloys including the medium/high-Mn TRIP/TWIP steels [15,16], TWIP HEAs ($Cr_{10}Mn_{40}Fe_{40}Co_{10}$ and CrMnFeCoNi) [12,17] and TRIP dual-phase HEAs ($Cr_{10}Mn_{30}Fe_{50}Co_{10}$) [9,10]. The correlation between the SFE and the deformation modes offers a practical design strategy through optimizing chemical composition and concentrations, providing that the SFE versus chemistry and concentration relations are precisely established [15,16]. However, currently such database in HEAs is still missing, and it is cumbersome to determine the SFE as a function of composition in multicomponent alloys.

Beyond the dependence of the SFE on the nominal composition, interestingly, there are some studies suggesting that the local chemical order and chemical fluctuation greatly affect the SFE and the γ-surface, and thus the deformation behavior and mechanical properties [18,19,20,21]. However, the role of local chemical order or chemical fluctuation and to which extend they can be tuned in the apparently random HEAs remains an open question awaiting for more experimental evidence [22]. In fact, recent experiments have demonstrated that after the solid-solutioning treatment the element distributions in the equiatomic CrCoNi and CrMnFeCoNi alloys were very homogenous [7,23].

Besides the effort of studying local chemical effect on the SFE and its influence on the deformation mechanisms, there are a lot of activities focusing on developing non-equimolar HEAs [9,12,24]. In the CrCoNi-based system, alloys composition may change from the Co-based alloys to Ni-based alloys, while keeping random solid solution with the fcc (and/or hcp) structures [25]. Already in this simple ternary case, there are nearly infinite ways to tune the composition, but a quantitative parameter to guide alloy design is still missing. Furthermore, depending on the nominal composition and temperature, an alloy can be ferromagnetic (FM) or paramagnetic (PM). However, the influence of magnetic state on the SFE (as well as the γ-surface) and the primary deformation mechanisms remains unclear. Previous theoretical studies mostly adopt the nonmagnetic (NM) approximation for the PM state [26-28], which totally neglects the effects of the local spins [29]. Additionally, if chemical fluctuations or short-range orders are present (e.g., accelerated by high-temperature aging on purpose [21]) and affect the SFE, one should also consider the magnetic effect that couples with the local chemical environment.

In the present work, using *ab initio* calculations we systematically studied the chemical and magnetic dependence of the γ-surface in the fcc $Cr_xCo_yNi_{100-x-y}$ ($10 \lesssim x, y \lesssim 40$, at.%) MEAs and explored their influence on the primary deformation mechanisms. Based on our theoretical results, we provided several design routes for tailoring the TRIP/TWIP effects in the Cr-Co-Ni alloys, which were further confirmed by our accordingly designed experiments. Specifically, we fabricated two Cr-Co-Ni alloys which demonstrate the desired TWIP and TRIP effects as well as excellent tensile properties. The present work demonstrates the predictive power of ab initio calculations in alloy design when physics-based principles are properly established and offers both guidelines and database for future studies of MEAs and HEAs.

The rest of the paper is organized as follows. In Section 2 details of the theoretical and experimental work are given. In Section 3, we put forward the concentration dependence of the magnetic states, the phase stability and the SFE of PM and FM Cr-Co-Ni alloys. Then we present the predicted plastic deformation modes based on the calculated effective energy barriers (EEBs) [30]. In Section 4 we propose the alloy design routes with desired deformation modes based on our theoretical results. In Section 5, experimental results are compared with the theoretical predictions. Conclusions are given in Section 6.

## 2. Methodology

### 2.1 Computational methods

First-principles calculations were performed by using the exact muffin-tin orbitals (EMTO) method based on density functional theory (DFT) [31-33]. The EMTO method is an improved screened Korringa-Kohn-Rostoker (KKR) method [34], where the one-electron potentials are represented by large overlapping muffin-tin potential spheres. Generalized gradient approximation (GGA) parameterized by Perdew, Burke and Ernzerhof (PBE) was used to treat the exchange-correlation density functional [35]. In the self-consistent calculations, the one-electron equations were solved within the soft-core scheme and scalar-relativistic approximation. The Green's function was calculated for 16 complex energy points distributed exponentially on a semi-circular contour, including states within 1 Ry below the Fermi level. The basis sets included *s*, *p*, *d*, and *f* orbitals. The Cr-$3p^64s^23d^4$, Co-$3p^64s^23d^7$, Ni-$3p^64s^23d^8$ were treated as valence states. The irreducible parts of the Brillouin zones were sampled by uniformly distributed *k*-points. The *k*-mesh was carefully tested and the 12×24×3 mesh was adopted for all calculations.

The chemical disorder was described with the coherent potential approximation (CPA) as implemented in the EMTO method [36]. CPA is an efficient and useful approach to treat the compositional and magnetic disorder in random solid-solution alloys. In the present study, we considered both FM and PM states. The PM state was simulated by the static disordered local moments (DLM) model [37], the possible thermal spin fluctuations were however neglected [38]. Since we only considered temperature effects up to the room temperature, in the free energy *F*(*T*,*V*) we included the effect due to thermal lattice expansion as well as the magnetic entropy term for the PM state [35]. The electronic entropy and lattice vibrational terms were neglected as their contributions to energy were estimated small at room temperature [29,39]. The equilibrium lattice parameters at the PM and FM states were obtained from the Morse-type of equation of state and the lattice parameters corresponding to temperature *T* were derived by using the rule of mixtures and experimental thermal lattice expansion coefficient of pure metals [40] according to $\alpha_{alloy} = 1/n \sum_i^n \alpha_i$, where *n* is the number of elements and $\alpha_i$ the thermal lattice expansion coefficient of element *i*. According to the previous study [40], the calculated thermal lattice expansion coefficients show a good agreement with the experimental values for the MEAs around room temperature. For the sake of comparison, the calculations for the PM and FM states were carried out adopting the same thermal expansion coefficient.

The generalized stacking fault energy surface (GSFE, γ-surface) was determined by a rigid shift of one part of the fcc structure along the 1/6<11-2> direction in the {111} slip planes following the method in Ref. [41]. The calculations were performed using a supercell containing 12 fcc {111} layers with the stacking sequence of ABCABCABCABC for the fcc structure. The intrinsic stacking fault energy (ISF, $\gamma_{isf}$), unstable stacking fault energy (USF, $\gamma_{usf}$) and unstable twining fault energy (UTF, $\gamma_{utf}$) were extracted from the GSFE curve.

## 2.2 Experimental methods

Three ingots with 40 mm in diameter were cast by the induction melting technique and then homogenized at 1200 °C for 6 hours followed by water quenching. All the ingots were then hot forged at 1000 °C to obtain plates with cross section of 20 mm by 20 mm. The $Cr_{36}Co_{36}Ni_{28}$ alloy was annealed at 950 °C for 40 minutes to obtain a mean grain size of 8 μm. The equiatomic CrCoNi alloy was annealed at 950 °C for 30 minutes to obtain a mean grain size of 8.3 μm. The $Cr_{30}Co_{30}Ni_{40}$ alloy was further cold rolled to 1 mm in thickness and then annealed at 950 ºC for 25 minutes to obtain a mean grain size of 8.7 μm. Accordingly, these three MEAs share the similar grain size, facilitating the understanding of fundamental deformation mechanisms by excluding the grain size effect. The grain size was measured by using the linear intercept method. Tensile specimens with gauge length of 10 mm, width of 5 mm and thickness of 1 mm were cut from the sheets. Tensile tests were conducted at an initial strain rate of 8.3 × $10^{-4}$ $s^{-1}$ at ambient temperature. The strain was measured by an extensometer until 0.3, then the extensometer was removed and the strain hereafter was represented by the displacement of the crosshead.

The microstructures of the MEAs before and after tensile tests were characterized by electron backscattering diffraction (EBSD), and electron channeling contrast (ECC) imaging techniques on a field emission scanning electron microscope (FE-SEM, LEO SUPRA 35). An accelerating voltage of 20 kV, a working distance of 15 mm and a step size of 100 nm were used to acquire the EBSD maps. The ECC images were obtained at an accelerating voltage of 20 kV and a working distance of 7 mm. The specimens for EBSD and ECC characterizations were prepared by electro-polishing in an ethanol solution of 10% perchloric acid at 25 V at room temperature. Deformation microstructures were also characterized by a transmission electron microscope operating at 200 kV (TEM, JEM-2100F), and the TEM foils were prepared using twin-jet electro polishing method by Tenupole-5 in a solution of 70% methanol, 20% glycerine and 10% perchloric acid with a voltage of 20 V at −20 °C. The phases of the MEAs were identified by X-ray diffraction (XRD) measurements on a Rigaku Smartlab X-ray diffractometer (Cu-Kα target; scanning range: 40˚-100˚; step size: 0.02˚; scanning rate: 1˚ $min^{-1}$).

# 3. Results

## 3.1 Magnetic phase diagram of Cr-Co-Ni alloys

In binary Ni-Cr alloys, the magnetic transition temperature (Curie temperature, $T_c$) decreases almost linearly from 627 K for pure Ni to zero for Cr concentration higher than ~13 at.% [42], while in binary Ni-Co alloys, $T_c$ increases linearly with Co addition [43]. This can be understood considering that Cr atoms prefer antiferromagnetic coupling with each other and with Ni atoms, where magnetic frustration leads to the reduction of magnetic moments and the decreased $T_c$ with increasing Cr concentration; whereas Co couples ferromagnetically with Ni, which causes the increase in $T_c$ with Co addition. In ternary alloys, the composition dependent $T_c$ may be estimated using a mean-field approach, $T_c = 2/3(E_{PM} - E_{FM})/k_B$, where $E_{PM} - E_{FM}$ is the total energy difference between the PM and FM states calculated at the 0K equilibrium state, and $k_B$ is the Boltzmann constant [44]. The obtained $T_c$ map as function of composition in Cr-Co-Ni alloys is shown in Fig. 1(a). One can find that $T_c$ decreases remarkably with increasing Cr and/or decreasing Co concentrations, in consistent with their effects in the binary alloys. In the low-Cr and high-Co corner ($x \lesssim 15, y \gtrsim 27$), $T_c$ is above room temperature (RT), indicating the preferred FM state at ambient conditions. The FM region extends towards the low-Co and high-Cr region when decreasing temperature, e.g., at 77 K. It is worth to note that the equiatomic CrCoNi MEA has a very low $T_c$ (<~30K), which is in line with the magnetic measurement showing that no magnetic order was found down to 2K [23]. To benchmark our $T_c$ results by the mean-field method, we compare the theoretical $T_c$ with the experimental values [45] for $Cr_xCo_{(100-x)/2}Ni_{(100-x)/2}$ ($20 \lesssim x \lesssim 30$) alloys in Fig. 1(b). It shows that we may overestimate $T_c$ by $\lesssim$50K at high Cr concentrations.

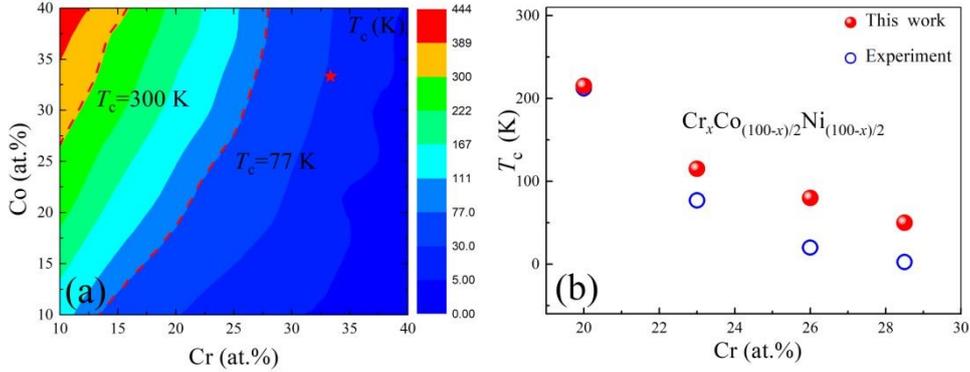

**Fig. 1** (Color online) (a) The composition dependent $T_c$ in the $Cr_xCo_yNi_{1-x-y}$ ($10 \lesssim x, y \lesssim 40$) alloys. The $T_c$ of 300K and 77K are marked as dash lines. The position of the equiatomic CrCoNi MEA is shown as a red star. (b) The calculated $T_c$ with respect to Cr concentration in the $Cr_xCo_{(100-x)/2}Ni_{(100-x)/2}$ ($20 \lesssim x \lesssim 30$) alloys, compared to the available experimental data [45].

## 3.2 Phase stability

In binary Ni-Co alloys, the experimental determined fcc/hcp phase boundary at room temperature locates approximately at ~33 at.% Ni with wide thermal hysteresis of the transformation process [43]. In binary Ni-Cr alloys, high Cr concentration promotes the formation of intermetallic $\sigma$ and body-centered cubic (bcc) phases [46]. For the ternary Cr-Co-Ni alloys, the equilibrium phase diagram shows that Cr-Co-Ni solid solutions are in the fcc structure for the whole composition range studied here at temperatures higher than 1027K [25]. The intermetallic $\sigma$ phase may occur for Cr concentration larger than ~50 at.% at high temperatures. Recent experiment has successfully obtained the single-phase fcc structure for the $Cr_{45}Co_{27.5}Ni_{27.5}$ alloy quenched from ~1400K, which is likely to be metastable at room temperature [47]. However, the equilibrium phase diagram for Cr-Co-Ni alloys at ambient and cryogenic temperatures is not completely established in the whole composition range, particularly for the phase stability relevant to the TRIP effect, i.e., fcc ($\gamma$) vs. bcc ($\alpha$) or fcc vs. hcp ($\varepsilon$) phases. In the following, we first assess the free energy differences ($\Delta F^{fcc-bcc} = F^{bcc} - F^{fcc}$ and $\Delta F^{hcp-bcc} = F^{bcc} - F^{hcp}$) between the $\alpha$, $\gamma$ and $\varepsilon$ phases at room temperature. We find that both $\Delta F^{fcc-bcc}$ and $\Delta F^{hcp-bcc}$ are positive for all compositions studied here, indicating that $\alpha$ phase is the least stable one. This is in line with previous experimental studies that no $\alpha$ phase has yet been found in the Cr-Co-Ni based MEAs and HEAs. Similarly, in the Co-rich entropic alloys (e.g., Co-(14-30)Cr-7.5Fe-7Mn-7.5Ni, at.%), the $\alpha$ phase was also found unstable by both thermodynamic and ab initio calculations, which was further confirmed by experiments [48]. Thereby, for the purpose of studying the deformation behavior of fcc Cr-Co-Ni alloys, the $\alpha$ phase becomes irrelevant and are not further considered in the following work.

In order to explore the composition dependent driving force for the structural transformation, we show the area-scaled free energy difference ($\Delta F^{fcc-hcp} = F^{hcp} - F^{fcc}$) between the hcp and fcc phases, $\gamma_0 \equiv 2\Delta F^{fcc-hcp}/A$, for both FM and PM states, in Fig. 2(a) and Fig. 2(b), respectively. According to the thermodynamic expression of the SFE [49], $\gamma_{isf} = \gamma_0 + 2\sigma$, where $\sigma$ is the fcc/hcp interfacial energy in the magnitude of several mJ/m$^2$ [50], $\gamma_0$ is usually considered as the first order approximation of $\gamma_{isf}$ and a direct indicator of the thermal stability of the fcc phase [51]. Therefore, negative $\gamma_0$ means that the $\varepsilon$ phase is energetically more stable than the $\gamma$ phase. For both magnetic states, $\gamma_0$ decreases with Co and Cr addition. Interestingly, Cr is found to be a stronger hcp stabilizer than Co, despite that pure Cr is a bcc metal while Co is an hcp metal at ambient conditions. At room temperature, the $\gamma$ phase is stable in the low-Co and low-Cr corner, and the $\varepsilon$ phase is stable in the high-Co and high-Cr region. It is important to notice that for low Co concentration, the $\gamma/\varepsilon$ phase boundaries (indicated by the solid line in Fig. 2(a-b)) at both FM and PM sates are similar; while for high Co concentration, the boundary is shifted toward the high Cr direction at the PM state compared to that at the FM one, which may have a similar magnetic origin as in pure Co [52,53]. The previous studies showed that the ground hcp state of Co is stabilized by magnetism at room temperature; and that phonon entropy, spin fluctuations and reduced magnetization all play important roles in the tendency to restore the fcc structure at high temperatures [52,53]. Actually, magnetism has been shown strongly influence the phase stability in the Cr-Co-Ni based HEAs [54, 55]. For instance, Niu et al. [54] showed that the hcp phase of the CrCoNi MEA is more stable than the fcc one at both magnetic and non-magnetic states; while for the CrMnFeCoNi HEA, the magnetically frustrated Mn atoms reduce the fcc-hcp energy difference, compared to the hypothetical NM case, which can be the reason for the absence of the hcp phase in CrMnFeCoNi HEA. Another similar work also indicated that the fcc and hcp phases compete with each other not only due to the volume effects but also due to the magnetism in a series of HEAs [55]. The present results and those in the above references highlight the importance of magnetism in studying the phase stability as well the deformation mechanisms (as discussed in the following sections) in alloys composed of 3d magnetic elements, e.g. Cr, Mn, Fe, Co and Ni.

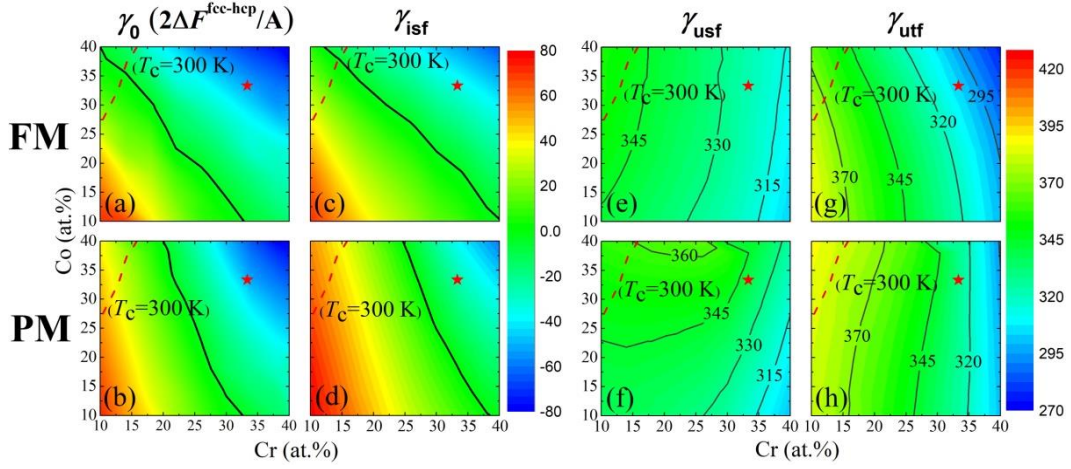

**Fig. 2** (Color online) The calculated $\gamma_0$, $\gamma_{isf}$, $\gamma_{usf}$ and $\gamma_{utf}$ for $Cr_xCo_yNi_{100-x-y}$ ($10 \lesssim x, y \lesssim 40$) alloys as a function of Cr and Co concentrations for FM (a,c,e,g) and PM states (b,d,f,h). Results are obtained at the room temperature. The $T_c$ of 300K is marked as the dash lines. The solid lines represent zero values of $\gamma_0$ and $\gamma_{isf}$ in (a-d). The position corresponding to the equiatomic CrCoNi MEA is indicated by a red star.

## 3.3 Stacking fault energy

The room-temperature $\gamma_{isf}$ maps are shown in Fig. 2(c) and (d) for the FM and PM states, respectively. They correlate nicely with the $\gamma_0$ maps at the corresponding magnetic state. Using the calculated data of $\gamma_0$ and $\gamma_{isf}$, the interfacial energy $\sigma$ for the present alloys is estimated to be in the range of 2~8.5 mJ/m$^2$ according to $\sigma = (\gamma_{isf} - \gamma_0)/2$, which agrees well with the previous theoretical values, e.g., 4~9 mJ/m$^2$ for CrCoNi MEA and CrMnFeCoNi HEA [56], 7.5~9 mJ/m$^2$ for Fe-Cr-Ni alloys [50]. It is worth noting that in the present alloys, $\sigma$ is always positive, indicating that it is possible to have small positive $\gamma_{isf}$ when the fcc structure is already thermodynamically unstable ($\gamma_0 < 0$). We mention that from structural point of view, $\sigma$ is the interfacial energy for the coherent interface composed of close-packed fcc{111} and hcp{0001} atomic planes, which is expected to be small and can have positive or negative values [50].

Here, we can quantitatively calculate the chemical dependence of the SFE, e.g., $\frac{\partial \gamma_{isf}}{\partial c_{Cr}}|_{c_{Co}}$ and $\frac{\partial \gamma_{isf}}{\partial c_{Co}}|_{c_{Cr}}$ at fixed Co or Cr concentrations (e.g., 10 at.%). At the FM state, the typical $\frac{\partial \gamma_{isf}}{\partial c_{Co}}|_{c_{Cr}}$ and $\frac{\partial \gamma_{isf}}{\partial c_{Cr}}|_{c_{Co}}$ are calculated to be approximately -2.0 mJ/m$^2$ per at.% Co and -2.4 mJ/m$^2$ per at.% Cr, respectively; while at the PM state, they are -1.2 and -3.2 mJ/m$^2$, respectively. The present trends are in accordance with the previously reported trends in both experimental and theoretical estimates [57-61]. Specifically, for the CrCoNi MEA, the calculated SFE is -24 mJ/m$^2$ for the PM state and -29 mJ/m$^2$ for the FM state, respectively. These values are consistent with the previous theoretical results at the fully random state, e.g., -18 mJ/m$^2$ [24], -24 mJ/m$^2$ [27] and -43 mJ/m$^2$ [18]. We mention here that one should not directly compare the theoretical SFE with the experimental values, e.g., 18±4 mJ/m$^2$ [62] and 22±4 mJ/m$^2$ [63]. The latter ones are usually calculated based on the measured partial dislocation separations ($d$) using the transmission electron microscopic (TEM) method according to $\gamma_{isf}^{exp} = \frac{\mu b^2}{8\pi d}\left(\frac{2-\nu}{1-\nu}\right)\left(1-\frac{2\nu \cos 2\alpha}{2-\nu}\right)$, where $\mu$ is the shear modulus in the {111} fault planes, $\nu$ is the Poisson ratio, $\alpha$ is the angle between the total Burgers vector and the dislocation line and $b$ is the magnitude of the partial Burgers vector [64]. Thereby, the experimentally estimated $\gamma_{isf}^{exp}$ is always positive as decided by the $\gamma_{isf}^{exp} \propto \frac{1}{d}$ relationship. Furthermore, in metastable alloys, in response to external stress, dislocations are easily separated into wide SF ribbons, even at very small strains [65], and therefore, it is not a trivial task to measure the partial separation $d$ [57], as compared to the cases of pure fcc metals. Smith et al. [66] measured the partial separation of the $a/2<110>\{111\}$ 60° mixed dislocations in CrMnFeCoNi HEA and found significantly large variations in $d$. Interestingly, one may notice that for the equiatomic CrCoNi MEA, the $\gamma_{isf}$ values obtained at various magnetic states (NM [26,27], PM and FM in the present work) do not differ significantly. It is likely because that the local magnetic moments on all atoms at the PM or FM states are very small or vanished. This observation may justify previously adopted NM approximation for the PM CrCoNi alloy at room temperature [26,27]. However, for the alloys with high-Co and low-Cr contents, $\gamma_{isf}$ differs significantly at different magnetic states, which highlights the importance to correctly consider the right magnetic state for the SFE calculation in various alloys and at different temperatures.

Fig. 2(e-h) show the maps of $\gamma_{usf}$ and $\gamma_{utf}$ for the two magnetic states. In contrast to strong variations in $\gamma_{isf}$, $\gamma_{usf}$ shows a much weaker chemical dependence. $\gamma_{utf}$ decreases strongly with increasing Cr concentration, but barely changes with Co addition for both magnetic states.

### 3.4 Plastic deformation modes

The three competing deformation modes in fcc materials are stacking fault (SF) formation, twinning, and full dislocation slip (SL). The activation of slip systems is known strongly dependent on the crystal orientation (Schmid's law). For example, deformation twinning is usually observed in grains orientated close to <111> direction in tensile test, while during compression twinning prefers to occur in grains with the <001> direction [67]. When we examine the favored slip mode on the primary slip plane, the preference of the activated deformation mode is governed by the combination of both Schmid's factor and the slip energy barrier. In the following, we adopt the concept of the effective energy barriers (EEBs) [30] which are defined as follows,

$$\bar{\gamma}_{sf}(\theta) = \frac{\gamma_{usf}}{\cos\theta}, \quad (1)$$

$$\bar{\gamma}_{tw}(\theta) = \frac{\gamma_{utf}-\gamma_{isf}}{\cos\theta}, \quad (2)$$

$$\bar{\gamma}_{sl}(\theta) = \frac{\gamma_{usf}-\gamma_{isf}}{\cos(60°-\theta)}, \quad (3)$$

where $\bar{\gamma}_{sf}$, $\bar{\gamma}_{tw}$ and $\bar{\gamma}_{sl}$ are the EEBs for SF, TW and SL, respectively. Here, the effect of crystal orientation on the activation of deformation modes is taken into account by $\theta$ (0°≤$\theta$≤60°), which is measured as the angle between the resolved shear stress and the twinning direction <11-2>. The preferred primary deformation mode is accordingly chosen by the lowest EEB. Note that within the current model, SF formation and hcp nucleation/growth (that is the deformation-induced martensitic transformation, DIMT) has equal EEBs due to the fact that the energy barriers for the partial dislocations to generate SF and hcp are the same. In other words, SF (or GSF) is a short-range type of planar defect in the normal direction of the fault, and therefore when two SFs locate on every other close-packed planes their interaction on the excess formation energy or the energy barrier is negligible. Using the calculated EEBs, one can construct a deformation mode map for the Cr-Co-Ni alloys as a function of composition. Here, we focus on the competition between SF and TW, which are independent of $\theta$, because the two modes are accomplished through slips of the partial dislocations of the same type. In contrast, SL is usually the favored mode at high $\theta$ when the tailing partial dislocation has large Schmid's factor than the leading/twinning partials [30]. We see here that even for the alloys with the largest $\gamma_{isf}$ (low-Co and low-Cr corner) in Fig. 2(d), TW is still predicted to occur at small $\theta$ by comparing $\bar{\gamma}_{tw}$ with $\bar{\gamma}_{sl}$, accompanied by SL at large $\theta$ values. Note that in polycrystals, the $\theta$ values in the primary/secondary slip planes in individual grains are generally different and they may also change during deformation due to grain rotation or intragranular lattice rotation [4].

Fig. 3(a-d) present the calculated $\bar{\gamma}_{sf}(0°) - \bar{\gamma}_{tw}(0°)$ as a function of Cr and Co concentrations at 300K and 77K for the FM and PM states, respectively. The GSFE maps at 77K appear similarly as those at 300K in Fig. 2 and are not shown here. It seems that the primary deformation mode is mainly decided by the $\gamma_{isf}$ when contrasting Fig. 2(c-d) with Fig. 3(a) and (c). In fact, when assuming the universal scaling law [68], namely, $\gamma_{utf} \approx \gamma_{usf} + 1/2\gamma_{isf}$, $\bar{\gamma}_{sf}(0°) - \bar{\gamma}_{tw}(0°)$ equals approximately $1/2\gamma_{isf}$, which may justify the above observation. For both magnetic states, positive $\bar{\gamma}_{sf}(0°) - \bar{\gamma}_{tw}(0°)$ locates at the low-Cr and low-Co alloys, indicating that these alloys are primarily deformed by TW. SF/MT is mostly preferred in the high-Co and high-Cr alloys. We have highlighted the ideal TW/SF boundary ($\bar{\gamma}_{sf}(0°) - \bar{\gamma}_{tw}(0°) = 0$) by thick lines in the maps. Decreasing temperature will shift the boundary to the low-Cr and low-Co corner, which promotes the fcc→hcp DIMT. Most interestingly, we see that the slopes of the boundary line in the FM and PM maps are quite different. We must emphasize here that the boundary should not be considered as rigid for several reasons. For example, the systematical errors in our calculations may change the position of the boundary. Additionally, Huang et al. [56] demonstrated that affine strain may also shift the boundary to the low SFE region. Furthermore, experiments usually show that DIMT and DT (deformation twins) are commonly mixed together in the form of nanoscale laths in the deformation microstructures in various metastable fcc alloys including the high-Mn TRIP and/or TWIP steels and the Cr-Co-Ni alloys [13,14,69], but the general trend is that the dominant deformation mode changes from DT to DIMT with decreasing the SFE, in addition to the dislocation slips. Therefore, one should not expect a sharp boundary separating the two deformation modes.

Bearing these facts in mind, for the well-characterized CrCoNi MEA, the primary deformation mode is predicted to be DIMT at room temperature, which is further promoted at 77K. This result is in line with the experimental observations at both room and cryogenic temperatures [4, 70]. For example, Miao et al. found that in the equiatomic CrCoNi MEA hcp lamellae appear at large strain levels and its volume fraction increases at cryogenic temperature [4]. However, the amount of hcp phase remains very low even at 53% true strain at cryogenic temperature, approximately, ~3%. While in the CrCoNi film fabricated using magnetron sputtering, Chen et al. [70] showed that the hcp phase that coexists with the fcc phase can reach a much higher fraction, ~60%. Both PM and FM calculations lead to the same result regarding the primary deformation modes due to the vanished magnetic moments. However, for the high-Co low-Cr alloys (e.g. the $Cr_{15}Co_{55}Ni_{30}$ alloy, $T_c$=385 K [13]), the predicted deformation modes are quite the opposite for the FM and PM states. Only when the magnetic state is properly accounted, one can predict the right deformation mode. In this case, FM $Cr_{15}Co_{55}Ni_{30}$ is predicted to deform by the DIMT, which agrees perfectly with the previous experimental study [13], while PM calculations would suggest TW.

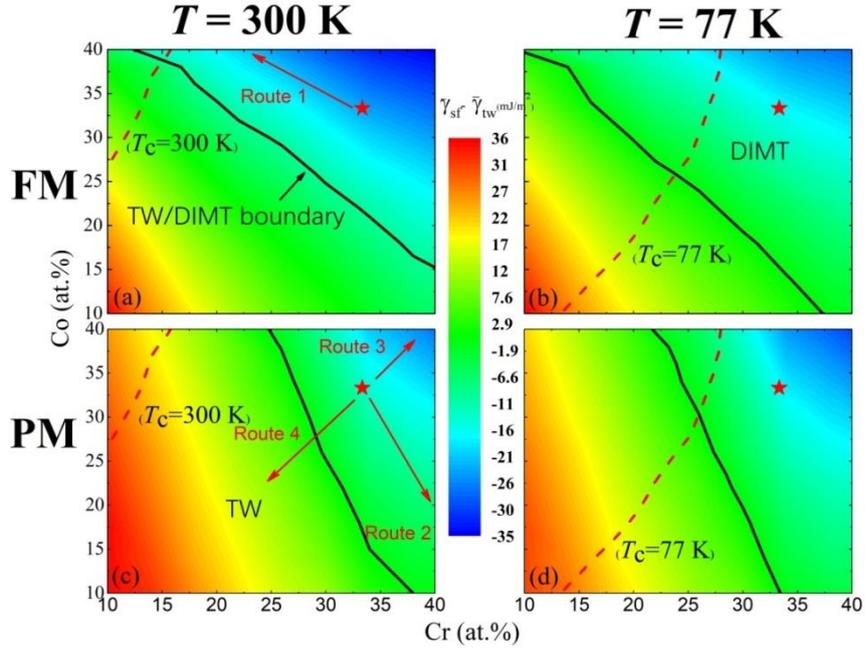

**Fig. 3** (Color online) The effective energy barrier difference between SF and TW, $(\bar{\gamma}_{sf}(0°) - \bar{\gamma}_{tw}(0°))$, for $Cr_xCo_yNi_{100-x-y}$, $(10 \lesssim x, y \lesssim 40)$ alloys at the FM (a) and PM (c) states (300K), respectively. The results at 77 K are shown in (b) and (d). The $T_c$ of 300K and 77K are marked as dash lines. The deformation mode boundary separating TW and DIMT is indicated by the solid lines. Starting from the equiatomic CrCoNi MEA, we suggest 4 designing routs based on the composition-deformation mode maps at the corresponding magnetic states.

## 4. Design routes

Based on the obtained composition and magnetism dependent maps of deformation modes (Fig. 3), one can design new alloys with desired deformation mechanisms. In the following, we start from a known alloy, e.g., the equiatomic CrCoNi MEA, and propose four routes for designing non-equiatomic TRIP/TWIP Cr-Co-Ni MEAs. This design strategy can minimize the effect of the systematical errors in our calculations and still take advantage of the predicted trends shown in the maps.

**(i) Route 1 towards FM high-Co low-Cr alloys;**

High Co concentration leads to high $T_c$, and these alloys are usually FM at room or cryogenic temperatures. As predicted in Fig. 3(a), for a fixed Cr concentration, the primary deformation mode changes from TW to DIMT with increasing Co concentration, which is in perfect agreement with the observations made by Remy and Pineau [13] in the $Cr_{15}Co_{75-x}Ni_x$ ($20 \lesssim x \lesssim 35$) alloys. They further showed that decreasing temperature can induce the deformation mode transition from DT to DIMT. In particular, when following a route approximately parallel to the TW/DIMT boundary in Fig. 3(a), it is predicted that the PM CrCoNi MEA and the FM $Cr_{15}Co_{50}Ni_{35}$ alloy (T4 alloy in Ref. [13]) should deform in a similar way, which is indeed confirmed by experiments [4,13]. Remarkably, these two alloys possess very similar tensile strength and ductility since they both deform by prevalent twinning and their twinnabilities are predicted similar [4,6,13]. Furthermore, from Fig. 3(a) we see that $Cr_{15}Co_{40}Ni_{45}$ (T6 alloy in Ref. [13]) is more stable than CrCoNi MEA, and accordingly its twinnability should be weaker. Indeed, it was observed that when $Cr_{15}Co_{40}Ni_{45}$ is deformed at 300K and 77K, twinning starts at around 30% and 10% strains, respectively; both are larger than the critical strains (12.9% at 300K and 6.7% at 77K) in CrCoNi MEA [63].

**(ii) Route 2 towards low-Co high-Cr alloys;**

High Cr and low Co concentrations quickly decrease $T_c$, therefore these alloys should be examined at the PM state (Fig. 3(c)). It is important to observe that the constant $\gamma_{isf}$ or $\bar{\gamma}_{sf}(0°) - \bar{\gamma}_{tw}(0°)$ lines at the PM state differ notably from those at the FM one. As an example, for the same Ni concentration, $Cr_{20}Co_{40}Ni_{40}$, $Cr_{30}Co_{30}Ni_{40}$ and $Cr_{40}Co_{20}Ni_{40}$ alloys have different $\gamma_{isf}$ values, thereby different twinnability at the PM state; while at the FM state they all have very similar $\gamma_{isf}$ values. Future experiments are needed to study the mechanical properties and deformation mechanisms in these alloys.

Furthermore, it was suggested that Cr atoms in the Cr-Co-Ni alloys lead to higher level of local lattice distortion and thereby cause higher friction stress against dislocation movement [47,71]. It is speculate that increasing Cr concentration while maintaining the TWIP or TRIP effects through adjusting Co concentration may lead to high yield strength and high ductility simultaneously. Recently, there are some experiments showing that when compared to the CrCoNi MEA, the $Cr_{40}Co_{20}Ni_{40}$ or $Cr_{45}Ni_{27.5}Co_{27.5}$ alloys show higher yield strengths while maintaining similar levels of ductility [47,72,73]. However, one should be aware of the fact that Cr decreases the SFE faster than Co, which can cause higher numbers of thermally induced twin boundaries in the microstructures prior plasticity test, as well as promote the early presence of SFs at small strains, these factors also contribute to a higher yield strength.

**(iii) Route 3 towards high-Co high-Cr alloys;**

Since both Cr and Co additions decrease the stability of the fcc solid solutions, the high-Co high-Cr alloys following Route 3 have even smaller $\gamma_{isf}$ and $\gamma_0$ compared to the CrCoNi MEA. Note that macroscopically the CrCoNi MEA deforms predominantly by DT and shows the typical TWIP effect [63]. By simultaneously increasing Co and Cr contents, the hcp phase quickly becomes more stable than the fcc one and we expect more DIMT and therefore the TRIP effect. Indeed, our experiments confirm this prediction, and the experimental details are presented in the following sections.

Furthermore, there may also exist the thermally induced hcp phase, realizing the hcp+fcc dual-phase microstructure in the as-quenched state when enough Co and Cr are added. As previously demonstrated in the Cr-Mn-Fe-Co alloys, with decreasing Mn concentration, i.e., decreasing the stability of the fcc structure and the $\gamma_{isf}$, the microstructure changes from the single fcc phase to the fcc+hcp dual-phase, which enables the TRIP-assisted dual-phase design in the $Cr_{10}Mn_{30}Fe_{50}Co_{10}$ HEA [10]. In the binary Co-Cr alloys with 85-70 at.% Co the area fraction of hcp in the quenched specimens was found more than 80% [48] and the martensitic start temperature ($M_s$) for the fcc→hcp phase transformation is about 750-1000°C [74]; while in the binary Co-Ni alloys, $M_s$ quickly decreases below the room temperature for the Ni concentration higher than ~30 at.% [43] and the martensitic transformation is remarkably sluggish [75,76]. Therefore, it awaits further experiments to find out at which composition/temperature the thermally induced hcp martensite occurs in the ternary alloys and its effect on the tensile properties.

In the Cr-Mn-Fe-Co-Ni quinary alloys, the hcp+fcc dual-phase structure has been successfully realized by increasing Co content and reducing the concentrations of the austenite stabilizing elements, e.g., Ni and Mn [48, 77]. For example, starting from the equiatomic CrMnFeCoNi HEA, Liu et al. [57] showed that increasing Co concentration to more than 27 at.%, compensated by the reduction in Ni content, leads to the occurrence of thermally induced hcp phase. Importantly, it results in a continuous improvement in both strength and ductility with increasing Co content by enhancing the TWIP effect and/or triggering the TRIP effect [57]. Similar design strategy has been adopted by Wang et al. [48] and Wei et al. [77] in this quinary HEA system, but with different concentration combinations, as well as in the quaternary Cr-Fe-Co-Ni alloys [78]. In principle, there are infinite ways to tailor the composition in the multiple component alloys, but the thermodynamic driving force should play a crucial role in affecting the fraction of the thermally induced hcp phase $f_{hcp}$. Wang et al. [48] argued that in the Cr-Mn-Fe-Co-Ni alloys, $f_{hcp}$ may be described by the Johnson-Mehl-Avrami-Kolmogorov type of expression in thermodynamic calculations, $f_{hcp} = 1 - exp\left[k\left(\frac{D_m}{RT}\right)^n\right]$, where $k$ and $n$ are fitting parameters, $D_m$ is the calculated driving force, $R$ is the gas constant and $T$ is the absolute temperature [48]. The above studies precisely emphasize the pressing demand of the theoretical studies as presented in the present work to save the experimental effort to seek for optimized alloys with desired phases and strengthening mechanisms.

**(iv) Route 4 towards low-Co low-Cr alloys;**

Opposite to Route 3, Route 4 points towards increasing the thermodynamical stability of the fcc structure through high Ni alloying. Increasing the $\gamma_{isf}$ will suppress DT, therefore following this route the alloys will deform more preferably by SL, towards the conventional Ni-based alloys. The experimental verification for this case is also presented in the following.

## 5. Experimental verification of alloy design

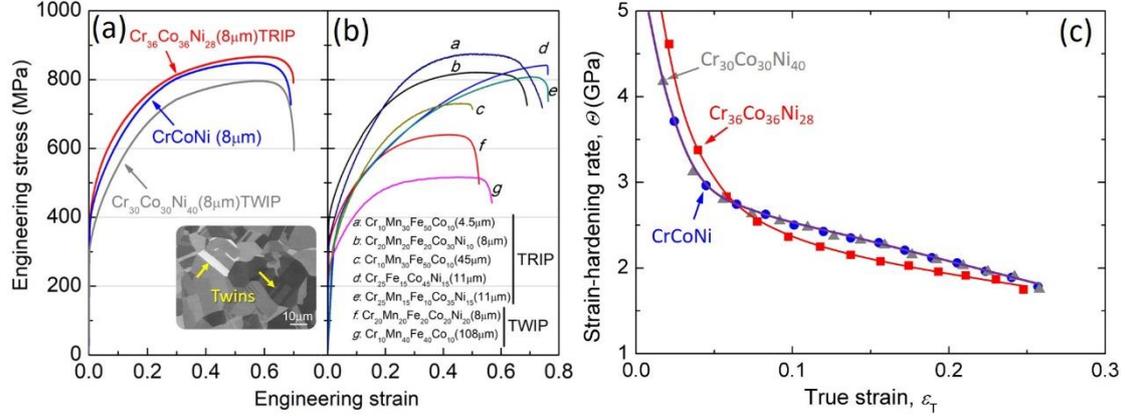

**Fig. 4** (Color online) Tensile engineering stress-strain curves of our TWIP and TRIP Cr-Co-Ni alloys (a), in comparison with other TWIP and TRIP MEAs and HEAs in literature [9,12,57,77,79] shown in (b). The insert figure in (a) is a typical recrystallized microstructure for $Cr_{36}Co_{36}Ni_{28}$ MEA. (c) Strain-hardening curves of our Cr-Co-Ni alloys in the initial stage of tensile tests.

The CrCoNi MEA has been extensively investigated due to the superior mechanical properties [3,4,63]. In this work, in addition to the equiatomic CrCoNi MEA as a reference alloy, other two MEAs ($Cr_{36}Co_{36}Ni_{28}$, $\gamma_{isf} = -38$ mJ/m$^2$ and $Cr_{30}Co_{30}Ni_{40}$, $\gamma_{isf} = -7$ mJ/m$^2$) were prepared according to the above proposed Routes 3 and 4. As designed, all three alloys exhibit a single phase fcc structure prior to deformation, and a typical SEM image of the $Cr_{36}Co_{36}Ni_{28}$ MEA is shown in the insert figure in Fig. 4(a). Equiaxed grains decorated with profuse growth twins were detected, indicating that these specimens were fully recrystallized. The mean grain size was determined to be about 8 μm for the three kinds of specimens, and in this case, the grain size effect on tensile properties and deformation mechanism can be excluded.

Mechanical properties were examined by tensile tests at room temperature, and the engineering stress-strain curves are shown in Fig. 4(a). It is interesting that elongations nearly approach the same value of 0.7 for the three kinds of MEAs albeit the relatively wide variation of compositions. While the $Cr_{30}Co_{30}Ni_{40}$ MEA exhibits the lowest flow stress among the three MEAs, the $Cr_{36}Co_{36}Ni_{28}$ and CrCoNi MEAs show very similar tensile curves. When compared to other TWIP and TRIP MEAs and HEAs (Fig. 4(b)), all three MEAs studied here possess extraordinary combination of strength and ductility. Furthermore, it is observed that once the TRIP effect (as confirmed in the following) is properly activated, it improves the strength and ductility of the alloys simultaneously, which clearly is not affected by the maximum configuration entropy.

Strain-hardening behaviors of three MEAs were examined and shown in Fig. 4(c). Overall, the three MEAs all possess excellent strain-hardening capability. While the $Cr_{30}Co_{30}Ni_{40}$ and CrCoNi MEA share the same strain-hardening curve, the $Cr_{36}Co_{36}Ni_{28}$ MEA with the lowest SFE shows higher strain-hardening rate when the true strain is smaller than 0.065, which can be induced by the early onset of profuse SFs acting as strong barriers for dislocation glide [80,81]. One more note to add is that the strain-hardening abilities caused by the mechanisms of $\gamma \rightarrow \varepsilon$ DIMT and DT (as evidenced in the following) are very similar in these MEAs, which calls for future studies to understand in detail that how dislocations interact with twin and fcc/hcp phase boundaries and their impact on the strain-hardening abilities.

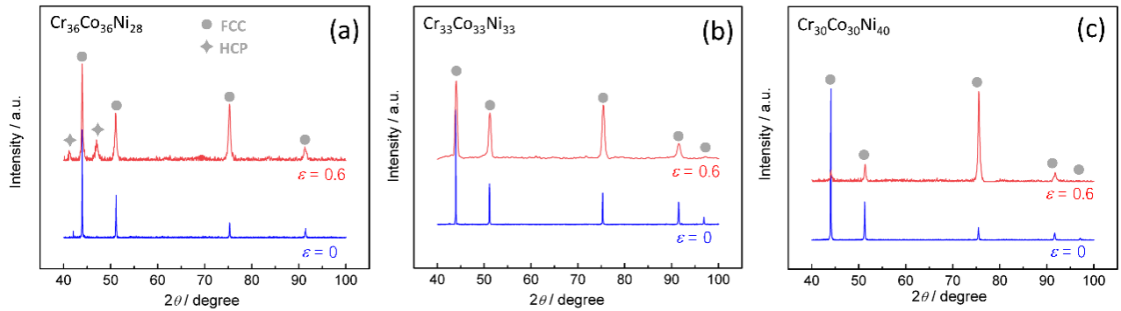

**Fig. 5** (Color online) XRD profiles of the three Cr-Co-Ni alloys before and after tensile tests.

In order to detect the phase constitution, the MEAs specimens before and after tensile tests were examined by XRD measurements (Fig. 5). After tensile tests, the uniform gauge area was precisely cut and examined. In contrast to the XRD profile of as-received specimen, no detectable peaks of new phase were observed in the well-studied CrCoNi MEA after tensile tests, as shown in Fig. 5(b). This is in line with the previous studies showing that the dominant deformation mechanisms are composed of planar dislocation slip and deformation twinning, accompanied by very limited number of nanoscale hcp lamellae at late stage of deformation [4]. For the $Cr_{30}Co_{30}Ni_{40}$ MEA as shown in Fig. 5(c), only fcc peaks were detected before and after tensile tests, which is in line with our prediction that both $\gamma_0$ and $\gamma_{isf}$ values are higher than those in CrCoNi MEA (Fig. 2(b) and (d)), which consequently inhibits any DIMT. In contrast, the $Cr_{36}Co_{36}Ni_{28}$ MEA with the lowest SFE shows apparent hcp peaks after tensile tests, indicating DIMT has been activated extensively in this newly designed $Cr_{36}Co_{36}Ni_{28}$ MEA.

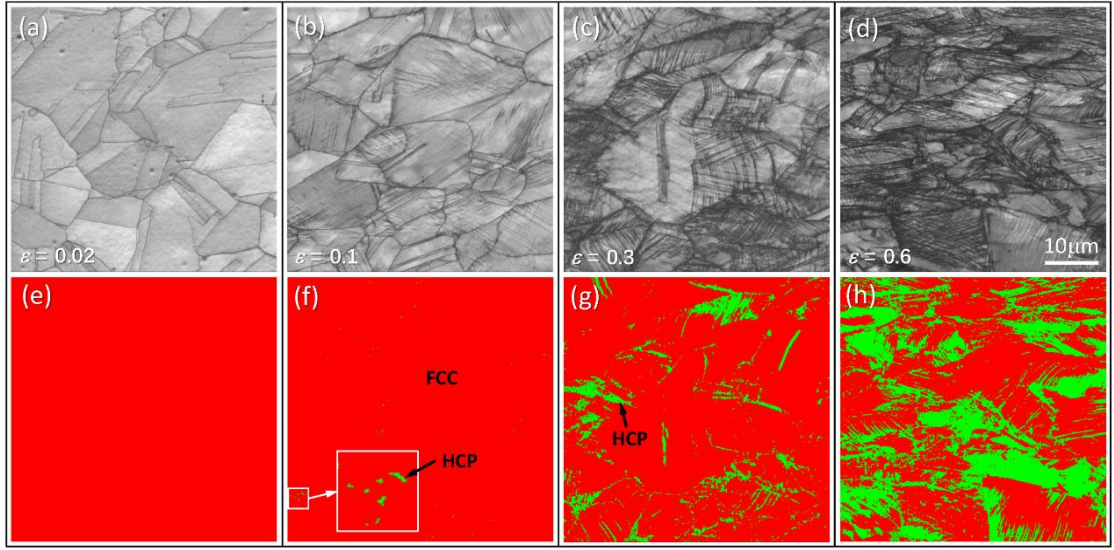

**Fig. 6** (Color online) EBSD images of deformation microstructures of the $Cr_{36}Co_{36}Ni_{28}$ alloy tensioned to specified strains. (a-d) image quality maps, (e-h) phase maps consisting of fcc and hcp phases.

The deformation microstructures of the $Cr_{36}Co_{36}Ni_{28}$ MEA at specified strains were further characterized by EBSD, as shown in Fig. 6. Deformation-induced hcp martensitic phase can be detected starting at a small strain of 0.1 and activates in large scale at a strain of 0.3, as shown in Fig. 6(f) and (g), which generally initiates along grain boundaries due to the large stress concentration. With increasing the strain to 0.6, the fraction of hcp martensitic phase becomes higher, as shown in Fig. 6(h). Inside the image quality maps in Fig. 6(a-d), some deformation traces in the form of straight lines can be detected besides the grain boundaries. These lines are related to the Kikuchi pattern quality which were formed due to the accumulation of planar dislocations, stacking faults, deformation twins or deformation-induced phase transformations.

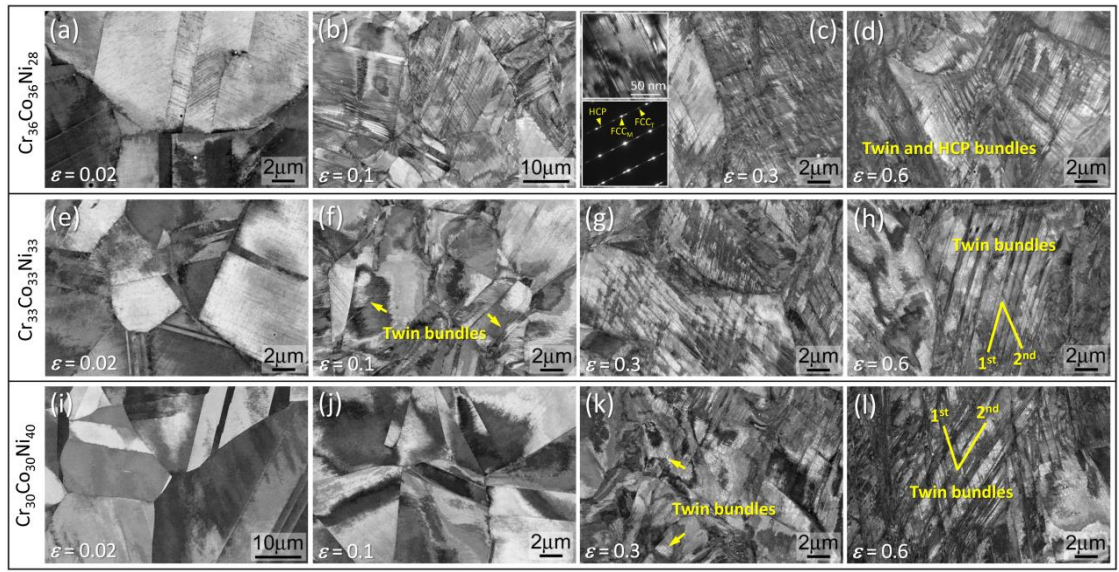

**Fig. 7** (Color online) Deformation microstructures at specified strains during tensile tests for the three Cr-Co-Ni alloys. (a-d) $Cr_{36}Co_{36}Ni_{28}$ alloy, (e-h) CrCoNi alloy, (i-l) $Cr_{30}Co_{30}Ni_{40}$ alloy.

The deformation microstructural evolution of three MEAs were also characterized by the SEM-ECC imaging technique at specified strains of 0.02, 0.1, 0.3 and 0.6, as shown in Fig. 7. Overall, all three MEAs are efficient in storing defects and sustaining high ductility. In the initial stage of deformation, higher density of planar dislocations and SFs are stored in the $Cr_{36}Co_{36}Ni_{28}$ MEA at the strain of 0.02, since dislocations are prone to dissociate and recovery is inhibited, leading to the higher strain-hardening rate as shown in Fig. 4(c). At the strain of 0.6, all the grains are full of dislocations and multiple deformation twin bundles and/or deformation-induced hcp martensitic phase. The hcp phase in $Cr_{36}Co_{36}Ni_{28}$ was further confirmed by TEM, as indicated by the bright filed image and the corresponding diffraction pattern in Fig. 7(c). Clearly, twinning is postponed to higher strains with increasing the Ni content (increasing the SFE, Fig. 2(d)). This microstructural evolution is fully consistent with our predicted deformation mode maps in Fig. 3(c).

The above experimental results indicate that the present modeling work is efficient and reliable in predicting the SFE and the corresponding deformation mechanisms. A new TRIP $Cr_{36}Co_{36}Ni_{28}$ MEA is developed which possesses even better mechanical performance than the well-known equiatomic CrCoNi MEA. The modeling work has provided ample room for alloy design, aiming for superior mechanical properties.

## 6. Conclusions

Using *ab initio* alloy theory, we investigated the chemical and magnetic effects on the deformation modes of the ternary Cr-Co-Ni alloys. We found that both Co and Cr additions lower the SFE and the fcc phase stability. Particularly, the Co- and Cr-rich alloys, compared to the well-studied equiatomic CrCoNi MEA, show even more negative $\gamma_0$ and $\gamma_{isf}$ values, indicating that the hcp phase is energetically more favorable than the fcc one and that the tendency of the occurrence of deformation-induced martensitic transformation (or the TRIP effect) is promoted by high Co and Cr concentrations. Our results also emphasize the critical role played by magnetism and show that the predicted deformation modes in Co-rich alloys are sensitive to the magnetic state. The obtained composition and magnetism dependent γ-surface enables us to establish maps of deformation modes, which are further used to guide designing TRIP and TWIP Cr-Co-Ni MEAs. We demonstrated that our theoretical predictions are consistent with available experimental observations in the literature. Furthermore, based on our *ab initio* calculations we designed both TWIP and TRIP MEAs which possess excellent combinations of strength and ductility. We believe that the present work will facilitate future alloy design, for instance, to further promote the TWIP and TRIP effects in this ternary alloy system by tailoring the composition and sheds lights on understanding the deformation mechanism in the Cr-Co-Ni based MEAs. This work will also facilitate the design of interstitial-alloyed (e.g., C and N) non-equiatomic Cr-Co-Ni MEAs in terms of the SFE and phase stability.

## Acknowledgement

The financial support is provided by the Major State Basic Research Development Program of China (2016YFB0701405). This work is also supported by the KTH-SJTU collaborative research and development seed grant in 2018, the Swedish Research Council, the Swedish Foundation for Strategic Research, the China Scholarship Council, the Swedish Foundation for International Cooperation in Research and Higher Education, and the Hungarian Scientific Research Fund (research project OTKA 128229), the Fundamental Research Funds for the Central Universities under grant No. N180204015. The computation resource provided by the Swedish National Infrastructure for Computing (SNIC) at the National Supercomputer Centre in Linköping is acknowledged.